\documentclass[prb, 10pt,aps, twocolumn,  notitlepage, showpacs, final,floatfix]{revtex4-1}
 
\pdfoutput=1
\usepackage{graphicx}
\usepackage{amsmath, amssymb, amsfonts, bm}
\usepackage{bbm}
\usepackage{latexsym}
\usepackage[colorlinks]{hyperref}
\usepackage{mathdots}
\usepackage[protrusion=true,expansion=true]{microtype}

\newcommand{\ee}{\mathrm{e}}
\newcommand{\ii}{\mathrm{i}}
\newcommand{\D}[1]{\,\text{d}#1\,}

\newcommand{\ket}[1]{\mathinner{|{#1}\rangle}}

\newcommand{\expes}[1]{\langle #1 \rangle}

\begin{document}
\title{Charge qubit driven via the Josephson nonlinearity}
\author{Jani Tuorila$^1$}
\author{Matti Silveri$^1$}
\author{Mika Sillanp\"a\"a$^{2,3}$}
\author{Erkki Thuneberg$^1$}
\author{Yuriy Makhlin$^{2,4,5}$}
\author{Pertti Hakonen$^2$}

\affiliation{$^1$Department of Physics, University of Oulu, FI-90014, Finland \\
$^2$O. V. Lounasmaa Laboratory, Aalto University, P.O. Box 15100, FI-00076 AALTO, Finland\\
$^3$Department of Applied Physics, Aalto University, P.O. Box 11100, FI-00076 AALTO, Finland\\
$^4$Landau Institute for Theoretical Physics, Kosygin st. 2, 119334, Moscow, Russia\\
$^5$Moscow Institute of Physics and Technology, 141700, Dolgoprudny, Russia}

\date{\today}

\begin{abstract}
We study the novel nonlinear phenomena that emerge in a charge qubit due to the interplay between a strong microwave flux drive and a periodic Josephson potential. We first analyze the system in terms of the linear Landau-Zener-St\"uckelberg model, and show its inadequacy in a periodic system with several Landau-Zener crossings within a drive period. Experimentally, we probe the quasienergy levels of the driven qubit with an $LC$-cavity, which requires the use of linear response theory. We also show that our numerical calculations are in good agreement with the experimental data.
\end{abstract}

\maketitle

\section{Introduction}

\begin{figure}[tH]
\centering
\includegraphics[width=\linewidth]{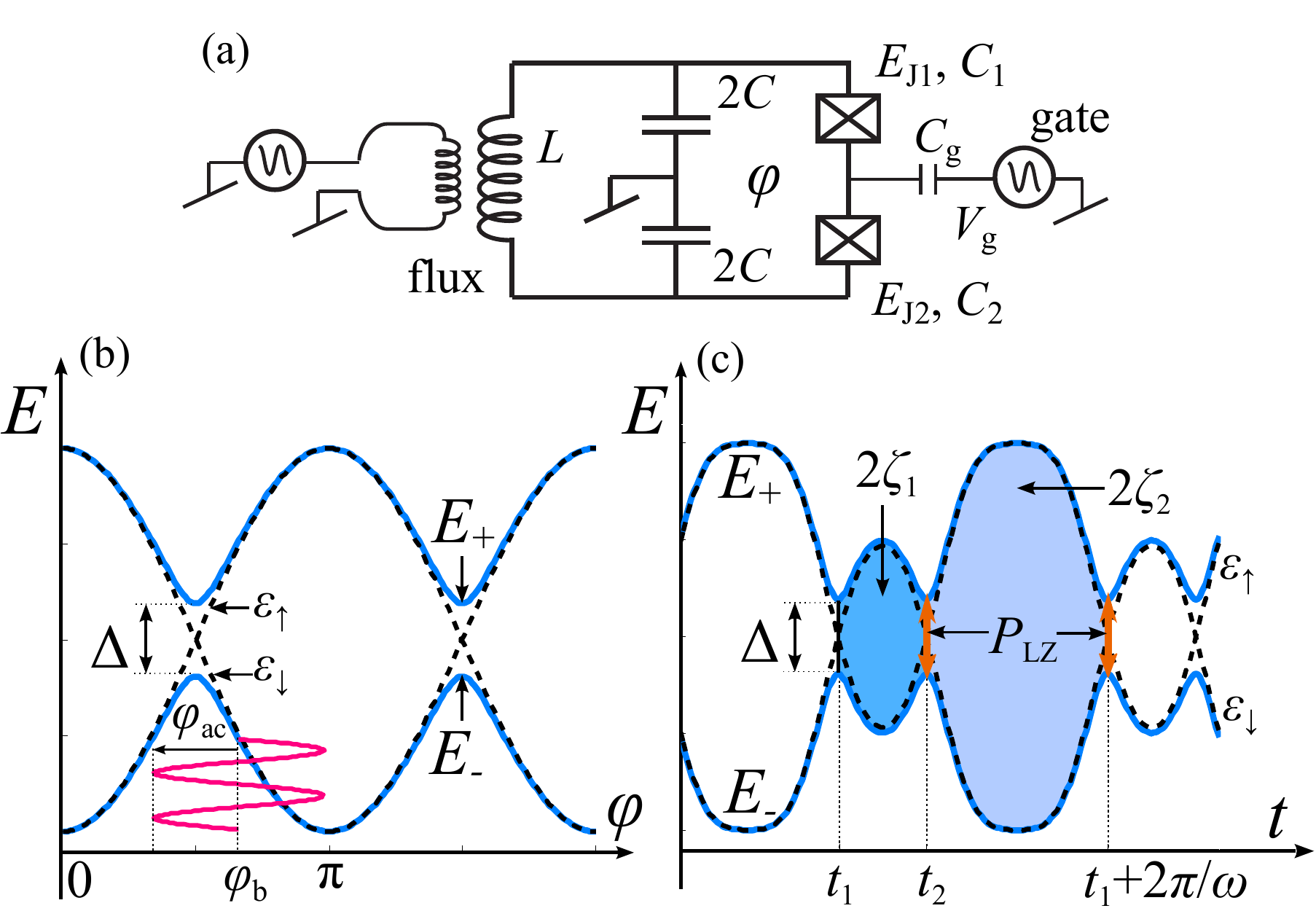}\\
\caption{Nonlinearly driven charge qubit. (a) Schematics of the experimental realization of an $LC$-cavity coupled inductively to a charge qubit driven by external flux $\varphi$. (b) Adiabatic qubit energies $E_{\pm}$ (solid lines) as a function of the bias flux $\varphi_{\rm b}$. Dashed lines indicate the diabatic energies $\varepsilon_{\downarrow \uparrow}$. The minimum gap is denoted with $\Delta$. The driving around the oscillation center $\varphi_{\rm b}$ is indicated with the red curve. (c) Qubit energies as a function of time. At an avoided crossing, the system can make an LZ-transition, which are indicated with orange arrows. The coloured areas indicate the phases $\zeta_{1,2}$ collected during the free adiabatic evolution in between LZ-transitions during one period of oscillation.}\label{fig:schema}
\end{figure}

Superconducting circuits have made possible the study of fundamental quantum physics with parameter values inaccessible to conventional atomic, molecular, or optical systems~\cite{Schoelkopf08}. The coupling between a superconducting two-level system, a qubit, with either zero-point~\cite{Chiorescu04, Wallraff04,Johansson06} or driven~\cite{Nakamura01,Oliver05,SillanpŠŠ06,Wilson07} microwave radiation has become a paradigm within the field. Physical understanding of such systems is of fundamental interest due to the possible application in quantum information processing. 

One example of such a novel system is a superconducting circuit that consists of a single-Cooper-pair transistor (SCPT), the qubit, inductively coupled to an $LC$-cavity, depicted in Fig.~\ref{fig:schema}(a). It has been used in several experiments to demonstrate, e.g., the Josephson inductance~\cite{SillanpŠŠ04}, the coupling between the vibrational and electric states of an artificial molecule~\cite{Gunnarsson08}, and exceptionally large Stark and generalized Bloch-Siegert shifts~\cite{Tuorila10} that emerge from strong external driving. In this paper, we will concentrate on the effects that are caused by the nonlinear coupling between the qubit and the magnetic flux, originating in the sinusoidal Josephson potential. 

We examine the nonlinear flux dependence by driving the qubit strongly with a periodic flux modulation $\varphi(t)=\varphi_{\rm b}+\varphi_{\rm ac}\sin\omega t$ through the superconducting loop. Here $\varphi_{\rm b}$, $\varphi_{\rm ac}$ and $\omega$ are referred to as the dc bias, the ac amplitude and the drive frequency of the flux, respectively. Previously, we have used the cavity to probe the quasienergy spectrum of the driven qubit~\cite{Tuorila10, Silveri13}, and studied also the influence of the chosen qubit basis on the apparent shifts in the locations of the multi-photon resonances of the qubit under strong drive~\cite{Silveri12}. Here, we first neglect the coupling with the cavity and discuss the effects of the strong drive in terms of interference occuring between multiple Landau-Zener (LZ) transitions~\cite{LZSM}, i.e. the Landau-Zener-St\"uckelberg (LZS) model~\cite{Shevchenko10}, and in terms of the quasienergies~\cite{Shirley65,Silveri13}. We emphasize that due to the periodicity of the qubit energy levels with respect to the flux, the LZS-model should be revised under ultra-strong driving, i.e. when two or more avoided crossings are traversed within a drive period, similarly as in a tripartite system consisting of a superconducting qubit and two microscopic two-level systems~\cite{Sun10}. Next, we treat the cavity as a small perturbation probing the quasienergies of the driven system. We give an explanation for the novel nonlinear features that appear in the spectrum of the system as the drive amplitude $\varphi_{\rm ac}$ is increased. Finally, we make a comparison with the measurement data obtained with the experimental realization of the setup. We omit the detailed explanations of the experimental aspects that can be found in our earlier publications~\cite{SillanpŠŠ04,Gunnarsson08,Tuorila10}.

\section{Josephson driven qubit}

We write the SCPT in the charge qubit limit resulting in the complete Hamiltonian of the coupled system:
\begin{align}\label{eq:hami}
\hat{H} = \hbar&\omega_{\rm c} \hat{a}^{\dag}\hat{a} + \frac{\varepsilon_0}{2}\hat{\sigma}_z-\frac{E_{\rm J}}{2}\Big\{\cos\big[\eta(\hat{a}^{\dag}+\hat{a})+\varphi(t)\big]\hat{\sigma}_x\nonumber\\
&-d\sin\big[\eta(\hat{a}^{\dag}+\hat{a})+\varphi(t)\big]\hat{\sigma}_y\Big\}.
\end{align}
We see that the scaled magnetic flux of the qubit comprises the cavity and external components, $\hat{\varphi}=\eta(\hat{a}^{\dag}+\hat{a})$ and $\varphi(t)=\pi\Phi(t)/\Phi_0$, respectively. In the above, we have defined $\varepsilon_0=4E_{\rm c}(1-2n_{\rm g})$ as the charging  energy difference between the states, $E_{\rm c}=e^2/2C_{\Sigma}$ as the charging energy of a single electron, $E_{\rm J}=E_{\rm J1}+E_{\rm J2}$ as the total Josephson energy, and $d=(E_{\rm J1}-E_{\rm J2})/E_{\rm J}$ as the asymmetry of the two junctions. Due to the sinusoidal dependence on flux, the coupling between the cavity and the qubit can be highly nonlinear even at the quantum limit, depending on the value of the parameter $\eta=\sqrt{\pi Z/R_{\rm K}}$, where the characteristic impedance of the cavity $Z=\sqrt{L/C}$ and $R_{\rm K}=h/e^2$ is the resistance quantum. However, in our experimental setup we have $L=410$ pH, $C=10$ pF, and thus $\eta \sim 0.03 \ll 1$. This means that the nonlinearities can be revealed only by driving the qubit strongly with an external signal. Also, in the limit of small number of cavity quanta, the cavity coupling can be treated as a small perturbation leading into the qubit (system) and probe Hamiltonians
\begin{align}
\hat{H}_{\rm S} &= - \frac{E_{\rm J}}{2}\big[\cos\varphi(t)\hat{\sigma}_x-d\sin\varphi(t)\hat{\sigma}_y\big]\label{eq:Hsys}\\
\hat{H}_{\rm P} &= \hbar\omega_{\rm c} \hat{a}^{\dag}\hat{a} + \hbar g (\hat{a}^{\dag} +\hat{a})\big[\sin\varphi(t)\hat{\sigma}_x+d\cos\varphi(t)\hat{\sigma}_y\big], \label{eq:Hprobe}
\end{align}
respectively. In the above, the linear coupling between the qubit and the cavity is given by $\hbar g=\eta E_{\rm J}/2$. Additionally, we bias the system into the charge degeneracy, $n_{\rm g}=0.5$, in order to minimize the charge noise and to highlight the nonlinearities in the system.

We first neglect the influence of the cavity probe~(\ref{eq:Hprobe}) and consider the strongly driven qubit Hamiltonian~(\ref{eq:Hsys}) by rotating into the $\hat{\sigma}_{\rm x}$ eigenbasis:
\begin{equation}
\hat{H}_{\rm S} = \frac{E_{\rm J}}{2}\big[\cos\varphi(t)\hat{\sigma}_z-d\sin\varphi(t)\hat{\sigma}_x\big].
\end{equation} 
The energy separation of the qubit can be controlled with the static flux bias $\varphi_{\rm b}$. However, the dependence is periodic due to the sinusoidal coupling. In the following, we choose the bias fluxes from the interval $\varphi_{\rm b}\in [\pi/2,\pi]$, and the other choices are obtained by employing the periodicity and the symmetry of the energies with respect to the bias flux. In the absence of driving, the uncoupled ($d=0$) qubit energies $\varepsilon_{\uparrow\downarrow}=\pm\frac12 E_{\rm J}\cos\varphi_{\rm b}$ are degenerate when $\varphi_{\rm b}=\pi/2$. We refer to these as diabatic energies, and depict them in Fig.~\ref{fig:schema} with dashed lines. The introduction of coupling lifts the degeneracy, resulting in adiabatic energies $E_{\pm} = \pm \frac12 E_{\rm J}\sqrt{\cos^2\varphi_{\rm b}+d^2\sin^2\varphi_{\rm b}}$ with an avoided crossing at the diabatic degeneracy (Fig.~\ref{fig:schema}, solid lines). The minimum gap is given by $\Delta=dE_{\rm J}$ at the diabatic degeneracy. Periodic modulation of the flux bias leads to oscillations in the adiabatic energy splitting $E(t)=E_+(t)-E_-(t)$ of the qubit, shown schematically in Fig.~\ref{fig:schema}(c).

\begin{figure*}[tH!]
\centering
\includegraphics[width=\linewidth]{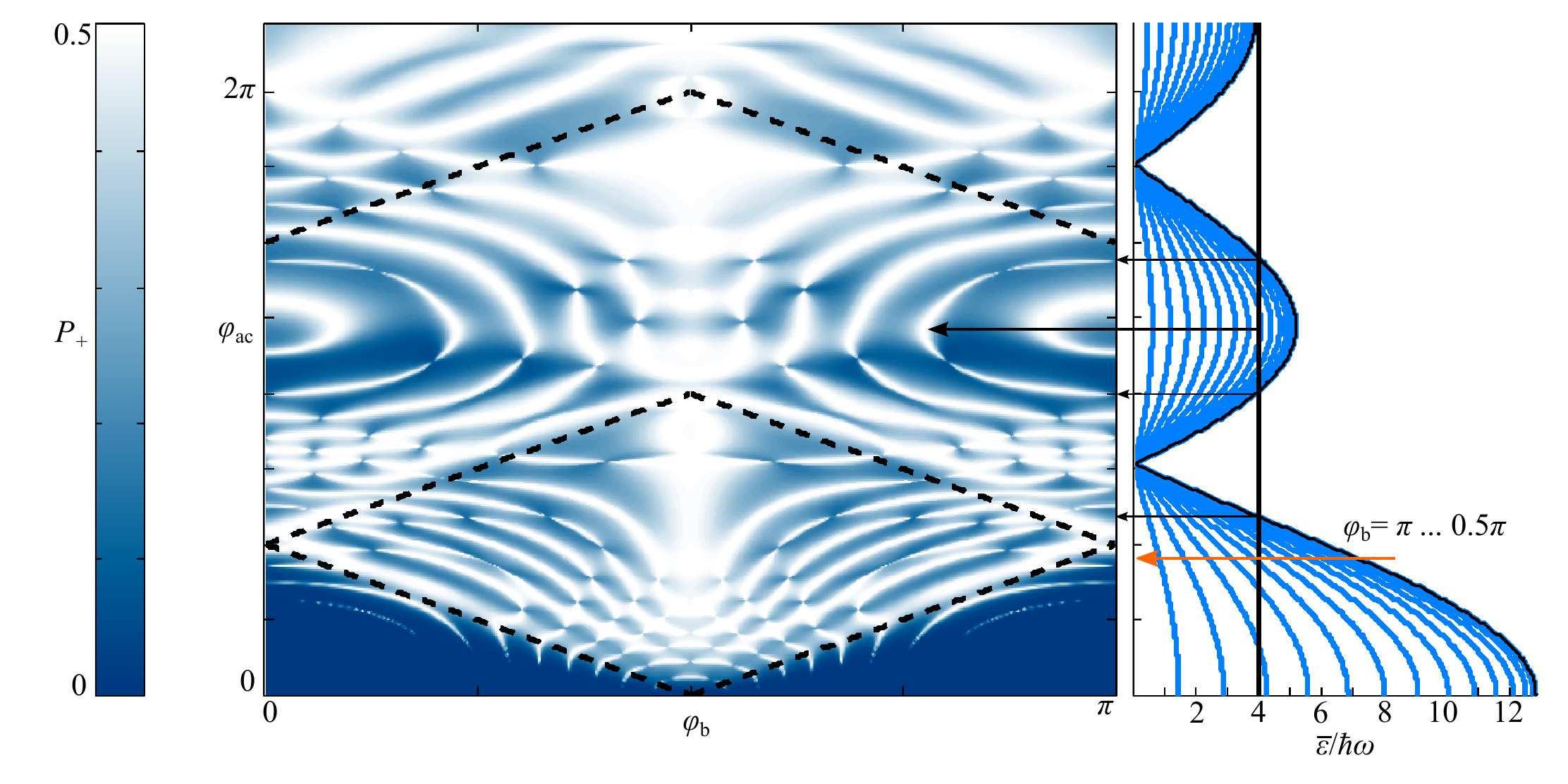}\\
\caption{Steady state population $P_+$ of the qubit's excited state in the $\varphi_{\rm b}$-$\varphi_{\rm ac}$ plane (left), together with a sketch of the geometric collapse (right). The blue curves indicate average energies $\bar{\varepsilon}$~\eqref{eq:geomc} with flux bias values ranging from $\pi/2$ to $\pi$, with the step $\pi/28$. As an example, we have denoted the location of the resonance with $m=4$ photons with the solid vertical line. Clearly, as one increases the drive amplitude $\varphi_{\rm ac}$ one also has to change the bias $\varphi_{\rm b}$ in order to stay on resonance. In certain regions of the amplitude, the average energy is below $4\hbar\omega$ with all values of $\varphi_{\rm b}$, and we say that the resonance is collapsed. After the first Bessel zero, the $m=4$ resonance is revived and then again collapsed. The extreme of the resonance is located at the first non-zero Bessel maximum. The locations of collapses and the revival of the $m=4$ resonance are indicated with horizontal arrows and are in accordance with a steady state maximum, which can be identified as the $m=4$ diabatic resonance~\eqref{eq:resodia} from Fig.~\ref{fig:reso}(a).}\label{fig:blochpop}
\end{figure*}

We have first solved the steady state population $P_+$ of the qubit's excited state as a function of the flux bias $\varphi_{\rm b}$ and the drive amplitude $\varphi_{\rm ac}$ by using the Bloch equations with relaxation times $T_1=15$ ns and $T_2 = 30$ ns. The parameter values $E_{\rm J}/h=27$ GHz, $d=0.19$,~\cite{Tuorila10} and $\omega/2\pi=2.1$ GHz are obtained from the experimental realization of the setup. We have neglected dephasing, and chosen values for $T_1$ and $T_2$ that are larger than in the experiments in order to resolve the individual resonances. The combined effect of driving and nonlinear coupling leads to rather elaborate interference pattern of  $P_+$, as shown in Fig.~\ref{fig:blochpop}. In the following, we will discuss the locations of resonances, and address the differences from the conventional LZS-model that arise from the periodic dependence on the control parameter.

\subsection{LZS-model}

Many previous experiments on strongly driven qubits~\cite{Oliver05, SillanpŠŠ06, Wilson07, Lahaye09, Li13} and other systems~\cite{Ditzhuijzen09,Stehlik12} have been discussed in terms of the Landau-Zener-St\"uckelberg model~\cite{Shevchenko10}. The heart of the model is the discretization of the dynamics into free adiabatic phase-evolution, interrupted with instantaneous Landau-Zener transitions at the locations of the avoided crossings. In our system, when $\varphi_{\rm ac}>\varphi_{\rm b}-\pi/2$, the flux bias reaches the minimum gap at least twice in the driving period. These values of the driving amplitude are located above the dashed line, starting at $(\varphi_{\rm b},\varphi_{\rm ac})=(\pi/2,0)$ in Fig.~\ref{fig:blochpop}. At the avoided crossing, the qubit can make an LZ-transition from one adiabatic state to another with the probability~\cite{LZSM}
\begin{equation}
P_{\rm LZ} = e^{-2\pi\delta},
\end{equation}
where $\delta=\Delta^2/4\hbar|\nu|$, and $\nu=\pm\omega E_{\rm J}\sqrt{\varphi_{\rm ac}^2-(\varphi_{\rm b}-\pi/2)^2}$ is the rate of change of the diabatic energy separation evaluated at the avoided crossing. We emphasize that the LZS-model in Ref.~\onlinecite{Shevchenko10} is linear in what comes to the dependence of the diabatic energies on the control parameter. However in our system, the energies are periodic due to the Josephson coupling (see Fig.~\ref{fig:schema}). When $\varphi_{\rm ac}>3\pi/2-\varphi_{\rm b}$, the system reaches the next avoided crossing and the corresponding LZ-processes should be included in the discretized dynamics in the LZS-model. Thus, the linear LZS-model should hold only within the dashed diamond indicated in Fig.~\ref{fig:blochpop}. Elsewhere the linear LZS-model cannot be expected to reproduce the $P_+$ calculated for the periodic qubit energy.

For now, we study amplitudes $\varphi_{\rm ac}$ in the validity range $[0,3\pi/2-\varphi_{\rm b}]$ of the linear LZS-model. First, we assume that the system is in the adiabatic ground state and study the excited state population after a single driving period. We call this the St\"uckelberg probability $P_{\rm St}$ to separate it from the population $P_+$ that is obtained by averaging over many periods. The cycle consists of two LZ-passages. During the cycle, the system collects two phases during the free adiabatic evolution (see Fig.~\ref{fig:schema}(c)):
\begin{equation}
\zeta_1=\frac{1}{2\hbar}\int_{t_1}^{t_2}E(t)\D t, \ \ \ \zeta_2=\frac{1}{2\hbar}\int_{t_2}^{t_1+2\pi/\omega}E(t)\D t.
\end{equation}
The system can evolve along two different paths which gives rise to interference, similar to a Mach-Zehnder interferometer. Starting the cycle at a time between $t_1$ and $t_2$, the important quantity is the adiabatic phase $\zeta_2$, accumulated in between the LZ-passages, that induces St\"uckelberg oscillations~\cite{Stuckelberg32} of the excited state population:
\begin{equation}\label{eq:stuck}
P_{\rm St} = 4P_{\rm LZ}(1-P_{\rm LZ})\sin^2(\zeta_2+\zeta_{\rm S}).
\end{equation}
Here, $\zeta_{\rm S}=\pi/4 + \delta(\ln \delta-1)+{\rm arg}[\Gamma(1-i\delta)]$ is the so-called Stokes phase~\cite{Kayanuma97}, collected during non-adiabatic transitions.

\begin{figure*}[tH!]
\centering
\includegraphics[width=\linewidth]{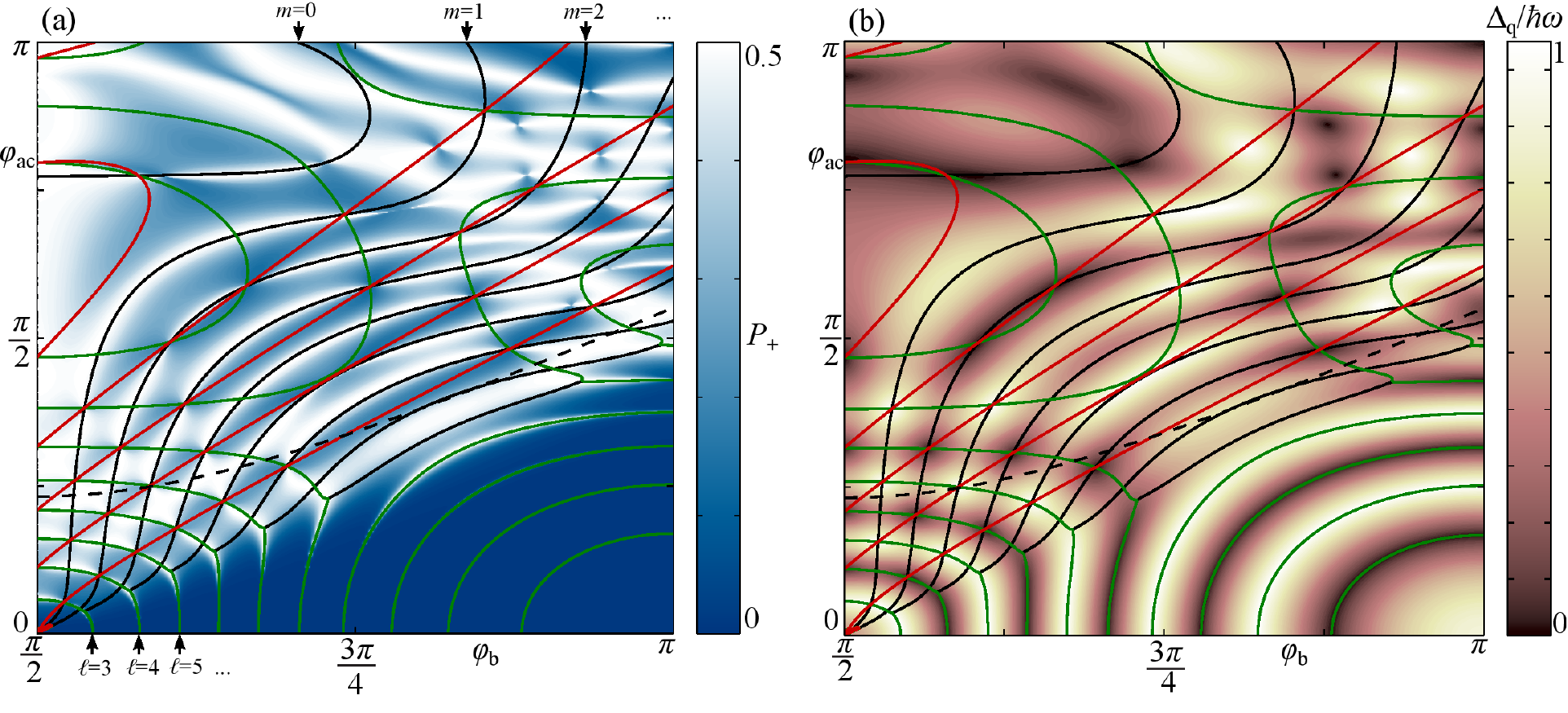}\\
\caption{(a) The excited state population $P_+$ from Fig.\ \ref{fig:blochpop} compared with the locations of the resonances from the linear LZS-model. The adiabatic resonance conditions (\ref{eq:resoadi}) are denoted by green lines and the 
diabatic conditions (\ref{eq:resodia}) by black lines. The first three visible adiabatic and diabatic resonances are marked with indices $\ell$ and $m$, respectively. Destructive St\"uckelberg interference (\ref{eq:deststu}) occurs on red lines, leading to CDT at the crossings with the resonances. Estimate for the adiabatic-to-diabatic behaviour is denoted with the dashed black line. (b) The continuous color scale shows the quasienergy difference $\Delta_{\rm q}$. The curves are the same as in panel (a).}\label{fig:reso}
\end{figure*}

Successive driving periods can also interfere, and by averaging over many periods we end up with the modulated St\"uckelberg, i.e. the steady state, population~\cite{Shevchenko10}
\begin{equation}\label{eq:difgrating}
P_+=\frac{P_{\rm St}}{2\sin^2 \xi},
\end{equation}
similar to a multi-pass Mach-Zehnder interferometer, or to a multi-slit diffraction grating. In the above, 
\begin{align}
\sin^2\xi &= P_{\rm St} + \big[(1-P_{\rm LZ})\sin \zeta_+ - P_{\rm LZ}\sin \zeta_-\big]^2\\
\zeta_+ &= \zeta_1+\zeta_2+2\zeta_{\rm S}-\pi\\
\zeta_- &= \zeta_1-\zeta_2.~\label{eq:diaphase}
\end{align}
We see that in order to observe maximum $P_+$ we need to minimize $\sin^2\xi $. This gives us the conditions for the phases $\zeta_+$ and $\zeta_-$:
\begin{align}
\zeta_+ &= \ell\times\pi\label{eq:resoadi}\\
\zeta_- &= m\times\pi\label{eq:resodia}
\end{align}
with integers $\ell$ and $m$. In addition, we should maximize $P_{\rm St}$  by choosing $\zeta_+-\zeta_-=2k\times \pi$ with integer $k$. However, the most clear features in the Bloch-equation based results in Fig.~\ref{fig:blochpop} are the points around which $P_+$ changes rapidly. These correspond to the $P_{\rm St}=0$ which is obtained by
\begin{equation}
\zeta_+-\zeta_-=(2k+1)\times\pi.\label{eq:deststu}
\end{equation}
The equations (\ref{eq:resoadi}) and (\ref{eq:resodia}) have a simple physical interpretation. When the driving is weak, $P_{\rm LZ}\ll 1$, and the system follows the adiabatic state, in accordance with the adiabatic theorem~\cite{LandauQM}. In this adiabatic limit, the locations of the resonances are determined by $\zeta_+$, which we will in the following refer to as the adiabatic phase. Similarly, when the driving is strong, $P_{\rm LZ}\sim 1$, and the bending of the energy levels at the avoided crossing is not important due to the strong tendency to tunnel between the adiabatic states. In other words the system follows the diabatic states and the diabatic phase $\zeta_-$ gives the relevant locations of resonances. We call this the diabatic limit. Moreover, by neglecting altogether the effects due to the asymmetry $d$, we end up with an analytic result 
\begin{align}
\zeta_-&\approx -\frac{\pi E_{\rm J}}{\hbar\omega}J_0(\varphi_{\rm ac})\cos\varphi_{\rm b}.\label{eq:multiph}
\end{align}
By combining with Eq.~(\ref{eq:resodia}), we obtain the multi-photon resonance condition $-E_{\rm J}J_0(\varphi_{\rm ac})\cos\varphi_{\rm b}=m\hbar\omega$.

We have plotted the adiabatic and diabatic resonance conditions in Fig.~\ref{fig:reso}(a). By properly choosing the parameters, we can see a shift from adiabatic to diabatic following simply by increasing the LZ-probability, e.g. by increasing the amplitude of the drive. A rough estimate for the location of the transition is given when the LZ-parameter $2\pi\delta=1$, denoted in Fig.~\ref{fig:reso}(a) with a dashed black line. In Fig.~\ref{fig:reso}(a), we have also plotted the locations~\eqref{eq:deststu} of the zeroes of the St\"uckelberg population. Those indicate destructive interference after each period, phenomenon often referred to as coherent destruction of tunneling~\cite{Grossmann91} (CDT). 

Overall, we see that within the validity region of the linear LZS-model, the locations of resonances in the steady state population $P_+$ follow nicely the prediction given by the 'stroboscopic' time-evolution. Also, the magnitudes of the multi-photon resonances are modulated, due to the single-period St\"uckelberg factor. The insufficiency of the linear LZS-model when $\varphi_{\rm ac}>3\pi/2-\varphi_{\rm b}$ is most clearly seen in the mismatch between the locations of CDT in the numerical excited state population $P_+$ and the crossing points of the diabatic resonances and the St\"uckelberg zeroes. Finally, we stress that the choice of the initial time of the first driving period is ambiguous. By starting the cycle at a time between $t_2$ and $t_1+2\pi/\omega$, one obtains relations~\eqref{eq:stuck}-\eqref{eq:diaphase} where the roles of the phases $\zeta_1$ and $\zeta_2$ are exchanged. Experimentally relevant steady state population is obtained by properly averaging over these two possibilities. Here, we put an emphasis only on the resonance conditions~\eqref{eq:resoadi} and~\eqref{eq:resodia} together with the locations of CDT, which are independent on the choice of initial time of the cycle.

The locations and magnitudes of the resonances are best given in the quasienergy formalism~\cite{Shirley65}, which works also outside the validity region of the linear LZS-model and provides an intuitive physical interpretation in terms of microwave dressed states of the qubit. To demonstrate this, we plot in Fig.~\ref{fig:reso}(b) the difference between the two quasienergies in a Brillouin zone of the driven qubit~\cite{Silveri13}. The coupling between the qubit and the driving field is the strongest at the locations of multiphoton resonances, which is clearly seen as the overlap between the LZS resonance conditions~\eqref{eq:resoadi} and~\eqref{eq:resodia}, and the valleys and ridges of the quasienergy landscape. 

Quasienergies of the driven qubit have been thoroughly investigated before~\cite{Tuorila10,Silveri13,Son09} and, thus, are not studied further in detail here. Instead, we concentrate on the novel effects due to the periodic coupling of the control parameter, e.g. the characteristic bending of the resonances away from the avoided crossing as the drive amplitude is increased. This will be discussed in terms of geometric collapse in the following section. Finally, we stress that with large drive amplitudes, the LZS-model should be revised to take into account the periodicity of the energy levels with respect to the control parameter. In addition, even though the LZS-model predicts well the locations of the population extrema of the qubit states, the inclusion of the probe Hamiltonian~(\ref{eq:Hprobe}) is difficult within the model. For this reason, the experimental results need to be analysed within the quasienergy formalism (see Sec.~\ref{sec:absor}). 

\subsection{Geometric collapse}

The characteristic bending of the multi-photon resonances at the diabatic limit can be understood by studying the mean energy of the qubit within a drive period. Consider an energy surface with a nonlinear dependence over a control parameter $\varphi$. When the parameter is varied periodically, the mean of the energy over an oscillation period is generally not given by the mean of the parameter, i.e. $\bar{\varepsilon}\neq \varepsilon(\bar{\varphi})$. This geometric shift in the effective energy separation can be seen as the rectification of the drive, allowing the control over the energy with the oscillation amplitude, in addition to the possible dc-controls.

In general, the mean of the energy has to be calculated numerically. Nevertheless,	we obtain a qualitative understanding by considering the diabatic limit where we can approximate $d=0$. Thus, the mean energy splitting of the qubit is analytic and given by
\begin{equation}\label{eq:geomc}
\bar{\varepsilon} = -E_{\rm J}J_0(\varphi_{\rm ac})\cos\varphi_{\rm b},
\end{equation} 
which is exactly the same term that turns out in the diabatic multi-photon resonance condition~(\ref{eq:multiph}). We see that on average, the energy difference between the two qubit levels vanishes at the zeroes of the Bessel function $J_0$, sketched in Fig.~\ref{fig:blochpop}(b). We refer to this as geometric collapse, as it has its origin in the special topology of our circuit. To give an example, we have also indicated the location of the multi-photon resonance with $m=4$. Clearly, as the amplitude $\varphi_{\rm ac}$ is increased, one needs to increase $\varphi_b$ in order to stay on resonance with $4\hbar\omega$. We notice that in the vicinity of the zeroes of $J_0$, the average energy $\bar{\varepsilon}<4\hbar\omega$, irrespective of the value of the bias flux $\varphi_{\rm b}$, indicating the disappearance of the resonances. The chosen resonance can be revived by further increasing the amplitude. Eventually, even the $m=1$ resonance will vanish due to the attenuation of $J_0$ with increasing $\varphi_{\rm ac}$. This discussion is in good qualitative agreement with steady state population $P_+$ in Fig.~\ref{fig:blochpop}(a), and also with the experimental results presented in Fig.~\ref{fig:LionHead}.

Effect similar to geometric collapse is observed in any periodic lattice under periodic temporal drive, e.g. with the quasienergy minibands in periodic superlattices that interact with intense far-infrared laser radiation~\cite{Holthaus92}. There the phenomenon is referred to as dynamical collapse. However, in superlattices the drive couples linearly to the position operator, whereas in our system the coupling is explicitly nonlinear both in the parameter $\varphi_{\rm b}$ and in the operator. Consequently, the dynamical collapse in our system does not depend on the frequency of the ac drive, contrary to the case with superlattices.


\section{Absorptive reflection measurement and probe
susceptibility}\label{sec:absor}

We use the reflection protocol~\cite{Pozar} to measure the quantum
circuit consisting of the strongly driven qubit perturbed weakly by the
cavity. The measured quantity is the reflection coefficient
$\Gamma(\omega_{\rm P})=V_{\rm out}(\omega_{\rm P})/V_{\rm
in}(\omega_{\rm P})$, where $V_{\rm in}$ is the incoming probe signal
sent at frequency $\omega_{\rm P}$ via a transmission line with
characteristic impedance $Z_0$, and $V_{\rm out}$ is the signal
reflected at the boundary of the transmission line and the studied
circuit with impedance $Z$. By using these notions, the reflection
coefficient can be also expressed as 
\begin{equation}
\Gamma=\frac{Z-Z_0}{Z+Z_0}. \label{Gamma}
\end{equation}
Without the strong driving, cavity-induced qubit excitations would not
be possible due to the large detuning $dE_{J0}/\hbar-\omega_{\rm c}\gg
1/T_2$, which places the system in the dispersive regime~\cite{SillanpŠŠ04}.

\begin{figure*}[tH!]
\centering
\includegraphics[width=1.0\linewidth]{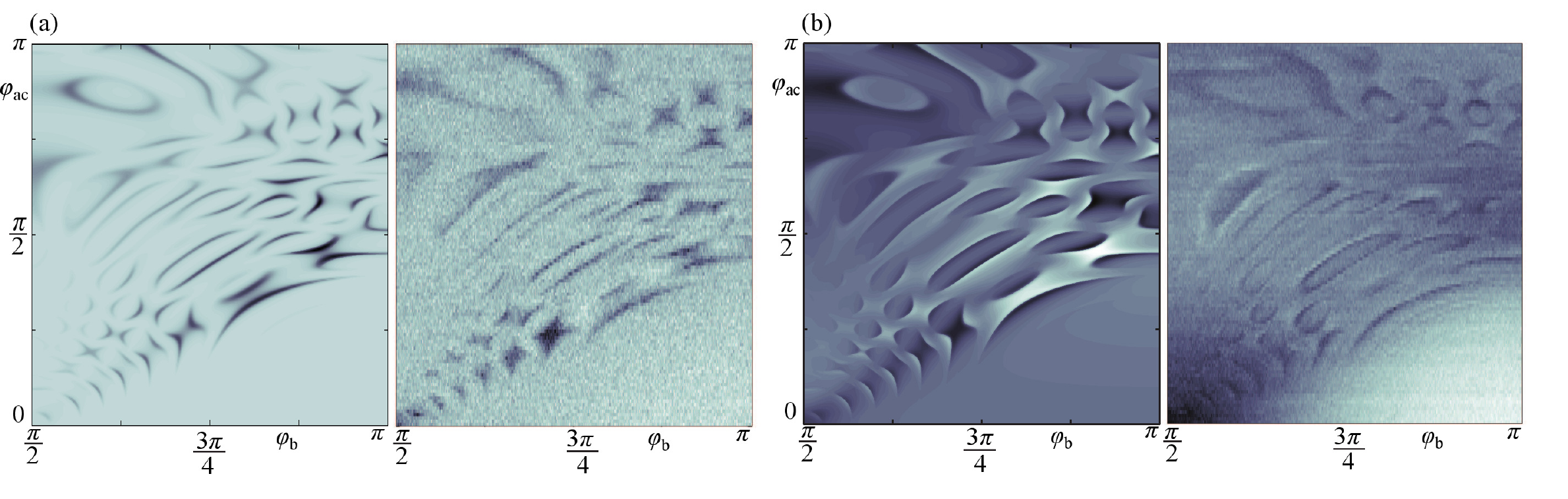}\\
\caption{Reflection from the $LCR$-oscillator, formed by the $LC$-cavity
and the strongly driven qubit. (a) Magnitude and (b) phase of the
reflection coefficient as a function of the flux bias $\varphi_{\rm b}$
and the drive amplitude $\varphi_{\rm ac}$. The dark indicates decreased
magnitude (phase). In both figures, we compare the simulation (left) to
the corresponding experimental result (right).}\label{fig:LionHead}
\end{figure*}

However, when the qubit is driven strongly, its energy bands deform
substantially into quasienergies, which energetically allow the cavity
induced excitations. By changing the flux bias $\varphi_{\rm b}$, we
observe a decrease in the probe reflection $|\Gamma|$ whenever the
cavity is on resonance with the driven qubit. We interpret this as the
energy flow~\cite{Tuorila09} from the cavity into the qubit, which has
to be compensated by an increased absorption (decreased reflection) from
the transmission line ($\omega_{\rm P}\approx\omega_{\rm c}$). At
resonance, energy difference between two quasienergy levels must be
equal with the energy quanta of the probe. With the experimental values
of $\omega/2\pi=2.1$~GHz and $\omega_{\rm P}/2\pi=3.47$~GHz, the
resonance condition can be written either as $\Delta_q=\hbar(\omega_{\rm
P}-\omega)$, or as $\Delta_q=\hbar(2\omega-\omega_{\rm P})$. Thus with a
fixed probe frequency, one can measure two distinct quasienergy
contours, as shown in Fig.~\ref{fig:LionHead}. We denote this type of
qubit measurement as the absorptive approach to distinguish it from the
more widely applied dispersive one (see, e.g., Ref.~\onlinecite{Blais04}).  

To accurately reproduce the measured reflection coefficient $\Gamma$~\eqref{Gamma}, the probe Hamiltonian~\eqref{eq:Hprobe} is considered in the semiclassical limit where quantum correlations between the cavity and the qubit are neglected. In this limit, the probe Hamiltonian~\eqref{eq:Hprobe} is written as 
\begin{equation}
\hat{H}_{\rm P}(t)=\varphi_{\rm c}(t) \frac{E_{\rm J}}{2} \left[\sin
\varphi(t)\hat{\sigma}_x+d\cos\varphi(t) \hat{\sigma}_y\right],
\label{weak.pert}
\end{equation}
where $\varphi_{\rm c}(t)=\varphi_{\rm c}(\ee^{-\ii \omega_{\rm
P}t}+\ee^{\ii \omega_{\rm P} t})$ and the constant energy of the cavity
is neglected. The probe Hamiltonian can also be expressed as 
\begin{equation}
 \hat{H}_{\rm P}(t)=\varphi_{\rm c}(t)\frac{\partial \hat{H}_{\rm
S}}{\partial \varphi}=-\varphi_{\rm c}(t) \hat{I}_{\rm q},
\label{weak.pert.i}
\end{equation}
where $\hat{I}_{\rm q}$ is the current operator for the qubit. In the
limit of $\eta\ll 1$, we have that $\varphi_{\rm c}\ll\varphi_{\rm ac}$
and the probe Hamiltonian can be considered as a weak harmonic
perturbation on the qubit. Then, the impedance of the qubit $Z_{\rm q}$
can be evaluated by applying the linear response theory~
\cite{LandauStat}. 

We first introduce the generalized probe susceptibility~
\cite{LandauStat} $\alpha(\omega_{\rm P})=\alpha'(\omega_{\rm P})+\ii
\alpha''(\omega_{\rm P})$, which determines the response
$\expes{\hat{I}_{\rm q}}(\omega_{\rm P})=\alpha(\omega_{\rm P})
\varphi_{\rm c}(\omega_{\rm P})$ to the probe Hamiltonian~\eqref{weak.pert.i}. The response implies the probe impedance $Z_{\rm
q}(\omega_{\rm P})=\ii\omega_{\rm P}\alpha^{-1}(\omega_{\rm P})$ of the
qubit. To calculate the generalized probe susceptibility $\alpha$, we
note that the absorption rate $\mathcal{P}(\omega_{\rm P})$
corresponding to $\hat{H}_{\rm P}$~\eqref{weak.pert.i} is proportional
to the imaginary part of the susceptibility $\alpha''(\omega_{\rm P})$
as $\mathcal{P}(\omega_{\rm P})=2\varphi_{\rm c}^2\alpha''(\omega_{\rm
P})/\hbar$~\cite{LandauStat}. On the other hand, the absorption rate
$\mathcal{P}(\omega_{\rm P})$ can be evaluated in the quasienergy basis
by applying Fermi's golden rule for quasienergy transitions~
\cite{Silveri13}:
\begin{equation}
 \mathcal{P}(\omega_{\rm P})=\frac{\varphi_{\rm
c}^2}{\hbar^2}\sum_{i,j}\frac{\gamma_{if}|\langle u_f|\hat{I}_q|u_i
\rangle|^2}{(\omega_{fi}-\omega_{\rm P})^2+\frac14 \gamma_{if}^2}.
\end{equation}
The matrix element between the final and initial quasienergy states
$\ket{u_{i,f}}$ is evaluated in the quasienergy basis,
$\omega_{fi}=(\epsilon_f-\epsilon_i)/\hbar$ (for absorptive transitions
$\omega_{fi}>0$) denotes the energy difference of the quasienergy
levels, and $\gamma_{ij}$ is the width (inverse lifetime) of the
transition. The Kramers-Kronig relations relate the real and the
imaginary parts of the probe susceptibility. 

Finally, the probe susceptibility $\alpha=\alpha'+\ii\alpha''$ and the
probe impedance $Z_{\rm q}=\ii\omega_{\rm P}\alpha^{-1}$ can be
calculated by exploiting the quasienergy structure and the probe
transitions. To relate this procedure to the reflection coefficient
$\Gamma$, we model the studied circuit as an $LCR$-oscillator with
impedance $Z_{\rm LCR}$ in parallel with the impedance $Z_{\rm q}$ of
the driven qubit. This as a whole is in series with the coupling
capacitance $C_{\rm c}$, producing the total impedance $Z$. The
resistance $R$ models the losses in the oscillator.  Within this
approach, the magnitude and the phase of the reflection coefficient are
calculated from Eq.~\eqref{Gamma} and plotted in the $\varphi_{\rm
b}$-$\varphi_{\rm ac}$ plane in Fig.~\ref{fig:LionHead} where we show
also a comparison with the experimental data with the same parameters. 

The calculated reflection coefficient shows a good agreement with the
experiments, both in the magnitude and the phase, as seen in Figs.~\ref{fig:LionHead}(a) and~\ref{fig:LionHead}(b), respectively. The
measured quasienergy contours display geometric collapse as they bend
towards larger $\varphi_{\rm b}$ when the drive amplitude $\varphi_{\rm
ac}$ is increased. This is the clearest indicator of the nonlinear
coupling of the drive, as explained by the LZS-model. Also, the
transition from the adiabatic to diabatic behaviour is visible, similar
to Fig.~\ref{fig:reso}(a).


\section{Conclusions}

We have presented a study of a charge qubit driven strongly via the Josephson potential. LZS-model shows good agreement with the full numerical Bloch solution of the excited state population. Outside the validity range of the linear LZS-model, we see deviations from the multi-photon resonance conditions, which we relate to additional evolutionary paths that lead to more complicated interference within a driving period. Thus, future work includes the revision of the LZS-model including the possibility of multiple LZ-crossings within a driving period. The absorptive response of the driven qubit to a weak probe cannot be calculated within the LZS-model. Instead, we have calculated the probe absorption spectrum of the quasienergies by employing the Fermi's golden rule. Absorption was shown to occur at the locations where the probe is on resonance with a quasienergy separation, and the resulting spectrum was in good accordance with the experimental data.

We have revealed the inherent nonlinearities within the system by driving with strong microwave flux, resulting in, e.g., the characteristic bending of the resonances and ultimately to the geometric collapse. Moreover, by enhancing the coupling between the cavity and the qubit in our circuit, we see a possibility of a novel study of nonlinear interactions at the quantum limit.

\acknowledgments

Fruitful discussions with Timo Hyart are gratefully acknowledged. This work was financially supported by the Finnish Academy of Science and Letters (Vilho, Yrj\"o and Kalle V\"ais\"al\"a Foundation), by the Academy of Finland, by the Cryohall infrastructure, by the European Research Council (grant No.~FP7-240387), and by the Academy of Finland (LTQ CoE grant no. 250280).


\begin{thebibliography}{99}
\bibitem{Schoelkopf08} R. J. Schoelkopf and S. M. Girvin, Nature \textbf{451}, 664 (2008).
\bibitem{Chiorescu04} I. Chiorescu, P. Bertet, K. Semba, Y. Nakamura, C. J. P. M. Harmans, and J. E. Mooij, Nature \textbf{431}, 153 (2004).
\bibitem{Wallraff04} A. Wallraff, D. I. Schuster, A. Blais, L. Frunzio, R.-S. Huang, J. Majer, S. Kumar, S. M. Girvin, and R. J. Schoelkopf, Nature \textbf{431}, 163 (2004).
\bibitem{Johansson06} J. Johansson, S. Saito, T. Meno, H. Nakano, M. Ueda, K. Semba, and H. Takayanagi, Phys. Rev. Lett. \textbf{96}, 127006 (2006).
\bibitem{Nakamura01} Y. Nakamura, Yu. A. Pashkin, and J. S. Tsai, Phys. Rev. Lett. \textbf{87}, 246601 (2001).
\bibitem{Oliver05} W. D. Oliver, Y. Yu, J. C. Lee, K. K. Berggren, L. S. Levitov, and T. P. Orlando, Science \textbf{310}, 1653 (2005). 
\bibitem{SillanpŠŠ06} M. Sillanp\"a\"a, T. Lehtinen, A. Paila, Yu. Makhlin, and P. Hakonen, Phys. Rev. Lett. \textbf{96}, 187002 (2006).
\bibitem{Wilson07} C. M. Wilson, T. Duty, F. Persson, M. Sandberg, G. Johansson, and P. Delsing, Phys. Rev. Lett. \textbf{98}, 257003 (2007).
\bibitem{SillanpŠŠ04} M. Sillanp\"a\"a, L. Roschier, and P. J. Hakonen, Phys. Rev. Lett. \textbf{93}, 066805 (2004).
\bibitem{Gunnarsson08} D. Gunnarsson, J. Tuorila, A. Paila, J. Sarkar, E. Thuneberg, Yu. Makhlin, and P. Hakonen, Phys. Rev. Lett. \textbf{101}, 256806 (2008).
\bibitem{Tuorila10}J. Tuorila, M. Silveri, M. Sillanp\"a\"a, E. Thuneberg, Yu. Makhlin, and P. Hakonen, Phys. Rev. Lett. \textbf{105}, 257003 (2010).
\bibitem{Silveri13} M. Silveri, J. Tuorila, M. Kemppainen, and E. Thuneberg, Phys. Rev. B \textbf{87}, 134505 (2013).
\bibitem{Silveri12} M. Silveri, J. Tuorila, M. Sillanp\"a\"a, E. Thuneberg, and P. Hakonen, J. Phys.: Conf. Ser. \textbf{400}, 042054 (2012).
\bibitem{LZSM}L. Landau, Phys. Z. Sowjet. \textbf{2}, 46 (1932); C. Zener, Proc. R. Soc. (Lond.) A \textbf{137}, 696 (1932).
\bibitem{Shevchenko10} S. N. Shevchenko, S. Ashhab, and F. Nori, Phys. Rep. \textbf{492}, 1 (2010).
\bibitem{Shirley65}J. H. Shirley, Phys. Rev. \textbf{138}, B979 (1965); Ya. B. Zeldovich, ZhETF \textbf{51}, 1492 (1966) [Sov. Phys. JETP \textbf{24}, 1006 (1967)].
\bibitem{Sun10} G. Sun, X. Wen, B. Mao, J. Chen, Y. Yu, P. Wu, and S. Han, Nat. Commun. \textbf{1}, 51 (2010).
\bibitem{Lahaye09} M. D. LaHaye, J. Suh, P. M. Echternach, K. C. Schwab, and M. L. Roukes, Nature \textbf{459}, 960 (2009).
\bibitem{Li13} Jian Li, M. P. Silveri, K. S. Kumar, J.-M. Pirkkalainen, A. Veps\"al\"ainen, W. C. Chien, J. Tuorila, M. A. Sillanp\"a\"a, P. J. Hakonen, E. V. Thuneberg, and G. S. Paraoanu, Nat. Commun. \textbf{4}, 1420 (2013).
\bibitem{Ditzhuijzen09} C. S. E. Ditzhuijzen, A. Tauschinsky, and H. B. van Linden van den Heuvell, Phys. Rev. B \textbf{80}, 063407 (2009).
\bibitem{Stehlik12} J. Stehlik, Y. Dovzhenko, J. R. Petta, J. R. Johansson, F. Nori, H. Lu, and A. C. Gossard, Phys. Rev. B \textbf{86}, 121303(R) (2012).
\bibitem{Stuckelberg32} E. C. G. St\"uckelberg, Helv. Phys. Acta \textbf{5}, 369 (1932).
\bibitem{Kayanuma97} Y. Kayanuma, Phys. Rev. A \textbf{55}, 2495 (1997).
\bibitem{LandauQM} L. D. Landau and E. M. Lifshitz, \textit{Quantum Mechanics: Non-relativistic Theory} (Pergamon, Oxford, 1965) pp. 140-142.
\bibitem{Grossmann91} F. Grossmann, T. Dittrich, P. Jung, and P. H\"anggi, Phys. Rev. Lett. \textbf{67}, 516 (1991).
\bibitem{Son09} S.-K. Son, S. Han, and S.-I. Chu, Phys. Rev. A \textbf{79}, 032301 (2009).
\bibitem{Holthaus92} M. Holthaus, Phys. Rev. Lett. \textbf{69}, 351 (1992).
\bibitem{Pozar} D. M. Pozar, \textit{Microwave Engineering} (John Wiley \& Sons, Inc. USA, 2005).
\bibitem{Tuorila09}J. Tuorila and E. Thuneberg, J. Phys.: Conf. Ser. \textbf{150}, 022092 (2009).
\bibitem{Blais04} A. Blais, R.-S. Huang, A. Wallraff, S. M. Girvin, and R. J. Schoelkopf, Phys. Rev. A \textbf{69}, 062320 (2004).
\bibitem{LandauStat}L. D. Landau and E. M. Lifshitz, \textit{Statistical Physics, Part 1} (Pergamon, Oxford, 1980).
\end{thebibliography}
\end{document}